\documentclass[doublecol]{epl2}

\title{Defect-induced ferromagnetism in graphite}
\shorttitle{Defect-induced ferromagnetism in graphite} 

\author{J. \v{C}ervenka \and C. F. J. Flipse}
\shortauthor{J. \v{C}ervenka and C. F. J. Flipse}

\institute{Physics Department, Eindhoven University of Technology - 5600 MB Eindhoven, The Netherlands}


\abstract{
We demonstrate direct evidence for ferromagnetic order at defect structures in highly oriented pyrolytic graphite with magnetic force microscopy at room temperature. Magnetic impurities have been excluded as the origin of the magnetic signal after careful analysis supporting an intrinsic magnetic behavior of carbon-based materials. The observed ferromagnetism has been attributed to originate from unpaired electron spins localized at grain boundaries. Scanning tunneling spectroscopy of grain boundaries showed intense localized states and enhanced charge density compared to bare graphite.
}

\begin{document}

\maketitle

\section{Introduction}
Graphite has been considered as a diamagnetic material for a~long time. However, recent experiments have shown that ferromagnetic order is possible in different carbon-based materials. Ferromagnetism with high Curie temperature, well above room temperature, and very small saturation magnetization has been reported in various graphitic systems \cite{Kopelevich,Coey,Esquinazi1,Esquinazi2,Esquinazi3,Mombru}. The~role of different magnetic impurities on the measured ferromagnetism has been studied in various samples of highly oriented graphite (HOPG), Kish graphite, and nature graphite \cite{Esquinazi2}. The~magnetization results showed no correlation with the~magnetic impurity concentration \cite{Esquinazi2}. Ferro- or ferrimagnetic ordering was demonstrated in proton-irradiated spots in highly oriented graphite \cite{Esquinazi1}. Bulk ferromagnetic graphite with a high defect concentration has been prepared via chemical route reaching the saturation magnetization 0.58 emu/g \cite{Mombru}. Apart from that, ferromagnetism has been observed in other carbon-based materials such as polymerized fullerenes \cite{Makarova}, carbon nanofoam \cite{Rode}, proton irradiated thin carbon films \cite{Ohldag} and nitrogen and carbon ion implanted nanodiamond \cite{Talapatra}. All these observations suggest an inherent ferromagnetic behavior of carbon-based materials.

Several theoretical investigations have been carried out to explain magnetism observed in these systems. The origin of ferromagnetism was suggested to be attributed to the mixture of carbon atoms with $sp^2$ and $sp^3$ bonds resulting in ferromagnetic interaction of spins separated by $sp^3$ centers \cite{Ovchinnikov}. Another theoretical calculation suggested magnetism in $sp^2$ bonded carbon nanostructures, which contains a negatively curved graphitic surface introduced via the presence of seven- or eight-membered rings \cite{Park}. In nanometer scale graphite, the electronic structure is strongly affected by the structure of the edges. Fujita and coworkers proposed that the $\pi$ electrons on a~monohydrogenated zigzag edge might create a ferrimagnetic spin structure on the edge \cite{Fujita}. Recently, it has been shown in spin-polarized density functional theory calculations that point defects in graphite such as vacancies and hydrogen-terminated vacancies are magnetic \cite{Lehtinen,Yazyev}. Three-dimensional network of single vacancies in graphite developed ferrimagnetic ordering up to 1~nm separation among the vacancies \cite{Faccio}.

In this Letter, we report an experimental study of ferromagnetic order in highly oriented pyrolytic graphite (HOPG) arising from defect structures. A~ferromagnetic signal has been observed locally with magnetic force microscopy (MFM) and in the bulk magnetization measurements using superconducting quantum interference device (SQUID). Observed ferromagnetism has been attributed to originate from itinerant electrons occupying narrow defect states of grain boundaries in the graphite crystal. A~systematic study of the grain boundaries have been performed on the same samples with scanning tunneling microscopy (STM) and spectroscopy (STS).

\section{Experimental}
Samples of HOPG of ZYH quality were purchased from NT-MDT. The ZYH quality of HOPG with the mosaic spread 3.5$^\circ$ - 5$^\circ$ has been chosen because it provides a~high population of step edges and grain boundaries on the graphite surface. HOPG samples were cleaved by an adhesive tape in air and transferred into a scanning tunneling microscope (Omicron LT STM) working under ultra high vacuum (UHV) condition. The HOPG samples have been heated to 500$^\circ$C in UHV before the STM experiments. STM measurements were performed at 78~K in the constant current mode with mechanically formed Pt/Ir tips. The same samples have been subsequently studied by atomic force microscopy (AFM), magnetic force microscopy (MFM) and electrostatic force microscopy (EFM) in air using Dimension 3100 SPM from Veeco Instruments. PPP-MFMR cantilevers made by NanoSensors with a hard magnetic material Co-coating film have been used in the MFM tapping/lift mode.

\section{Results and discussion}
\begin{figure}[t] 
   \centering
   \includegraphics[width=3.3in]{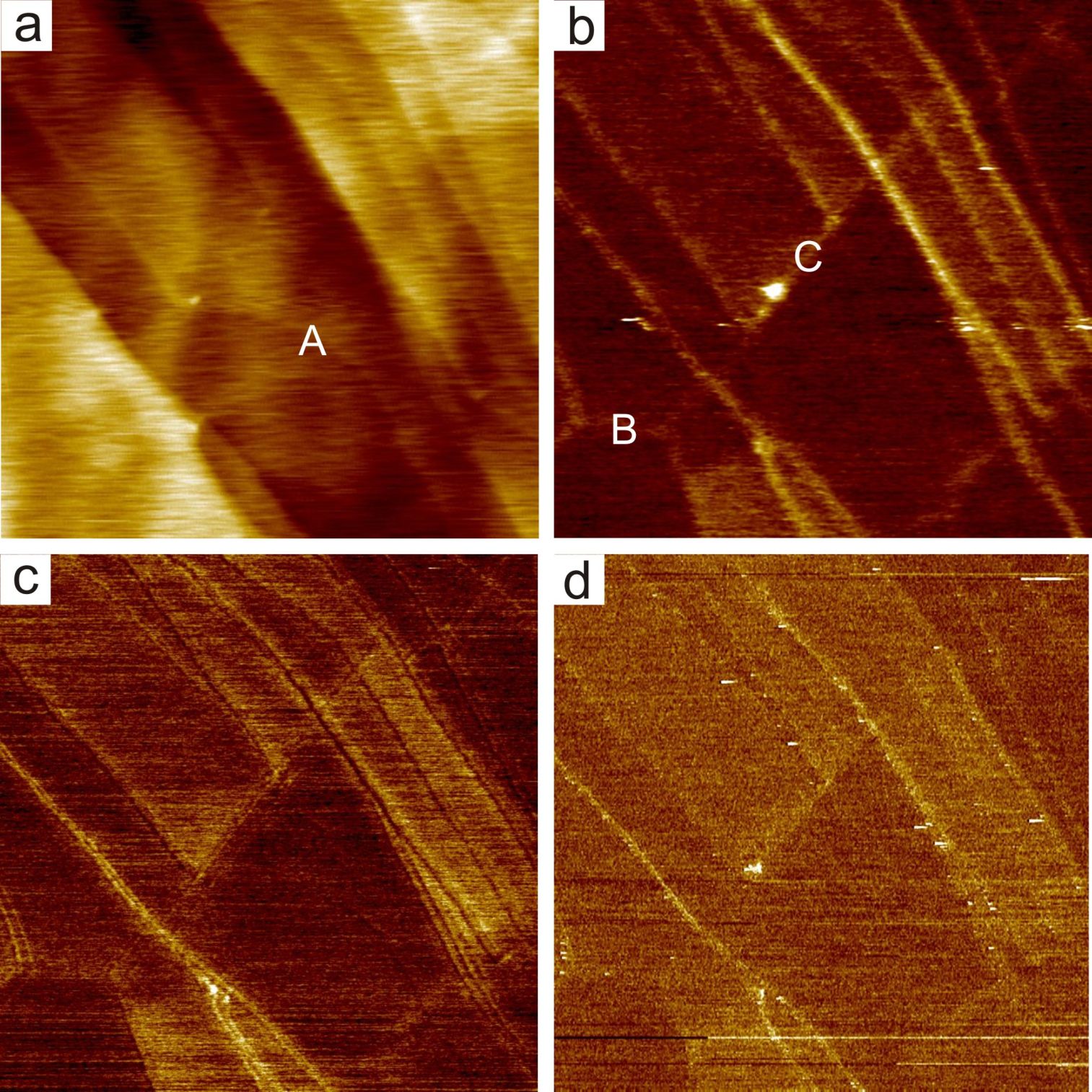}
   \caption{(Color online) The same area on the HOPG surface imaged with AFM (a), MFM (b) and (c), and EFM (d). MFM tip has been magnetized into the graphite surface (b) and out of the graphite surface (c), respectively. Image parameters: san area $2\times2~\mu$m$^2$, AFM $z$-range $z=5$~nm, MFM $z$-range (b) $\Phi = 2^\circ$ and (c) $\Phi = 1^\circ$, the MFM lift height $h = 50$~nm, EFM $z$-range $\Phi = 1^\circ$, the EFM lift height $h = 20$~nm.}
   \label{fig1}
\end{figure}

In figure~\ref{fig1}, AFM, MFM and EFM images of the same area on the HOPG surface are shown. AFM topography picture in figure~\ref{fig1}a displays a~surface with a~high population of step edges, surface distortions and defects. The MFM images in figure~\ref{fig1}b and figure~\ref{fig1}c were taken on the same place as the AFM image with a~lift scan height of 50~nm, where long-range van der Waals forces are negligible and magnetic forces prevail. Magnetic signal is measured on most of the line defects, however, a~step edge marked as A in figure~\ref{fig1}a does not show a magnetic signal in the MFM image. On the other hand, two lines in the MFM image in figure~\ref{fig1}b that are indicated as B and C do not show a noticeable height difference in the topography. The lines B and C are grain boundaries of HOPG. A detailed AFM and STM study can be found elsewhere~\cite{Cervenka1,Cervenka2}.

In order to determine the character of the detected magnetic signal, the MFM tip has been magnetized in two opposite directions: aiming into (figure~\ref{fig1}b) and out of the graphite surface plane (figure~\ref{fig1}c). Since the MFM signal represents the phase shift between the probe oscillation and the driving signal due to a magnetic force acting on the tip, the dependence of the phase shift on the force gradient can be expressed by a simple form \cite{Kebe}
\begin{equation}
\label{phase}
\Delta\Phi \approx -\frac{Q}{k}\frac{\partial F}{\partial z},
\end{equation}
where $Q$ is quality factor and $k$ is spring constant of the cantilever. Typical values of our MFM system give a minimal detectable force gradient in the order of 100~$\mu$N/m, $Q=200$ and $k=2.8$ N/m. For a true quantitative interpretation of MFM images it is necessary to have an exact knowledge of the geometry and magnetic properties of the tip and the substrate in order to express the force acting on the tip, which is difficult and has been achieved only in special cases \cite{Kebe}. Nevertheless, a~qualitative analysis can be done according to expression \ref{phase}, where a positive phase shift (bright contrast) represents a repulsive force between the tip and the sample, and a~negative phase shift (dark contrast) manifests an attractive interaction relative to the background signal. Since the tip magnetized into the graphite surface plane has shown a bright contrast in figure figure~\ref{fig1}b and out of plane magnetized tip produced a dark phase contrast on the line defects in figure~\ref{fig1}b, the orientation of the net magnetic moment in the defects stayed in the same direction, pointing out of the graphite surface plane. This shows a clear indication of ferromagnetic order at the defect sites at room temperature. In~the case of paramagnetic order, a bright contrast would be detected in both direction of the magnetization of the tip because the local magnetic moments would align with the magnetic field of the tip leading to attractive interaction. The same result would be valid if electric force gradients were detected due to charge accumulation at the step edges. Therefore, the ferromagnetic order in the defects of the HOPG sample is the only plausible explanation for the detected MFM signal.

However, not all the signal measured in the MFM showed to be sensitive to the reversal of the tip magnetization, in particular, areas with a different phase contrast. This is due to the metallic character of the magnetic coating film of the MFM tip, which probes electrostatic forces as well. Therefore EFM has been measured on the same place with Pt coated Si tip with a~lift scan height of 20~nm (see figure~\ref{fig1}d). A bright contrast is observed on the same places as in the MFM images. Similar observations of regions with a different potential has been measured in EFM and kelvin probe microscopy (KPM) on HOPG before \cite{Lu,Martinez-Martin}. This non-uniform potential distribution has been found to be caused by the mechanical stress induced during sample cleaving \cite{Martinez-Martin}. Thereby the MFM measurements represent a superposition of magnetic and electrostatic signal, which explains well the observed line shapes in figure~\ref{fig1}c.

\begin{figure}[t] 
   \centering
   \includegraphics[width=2.4in]{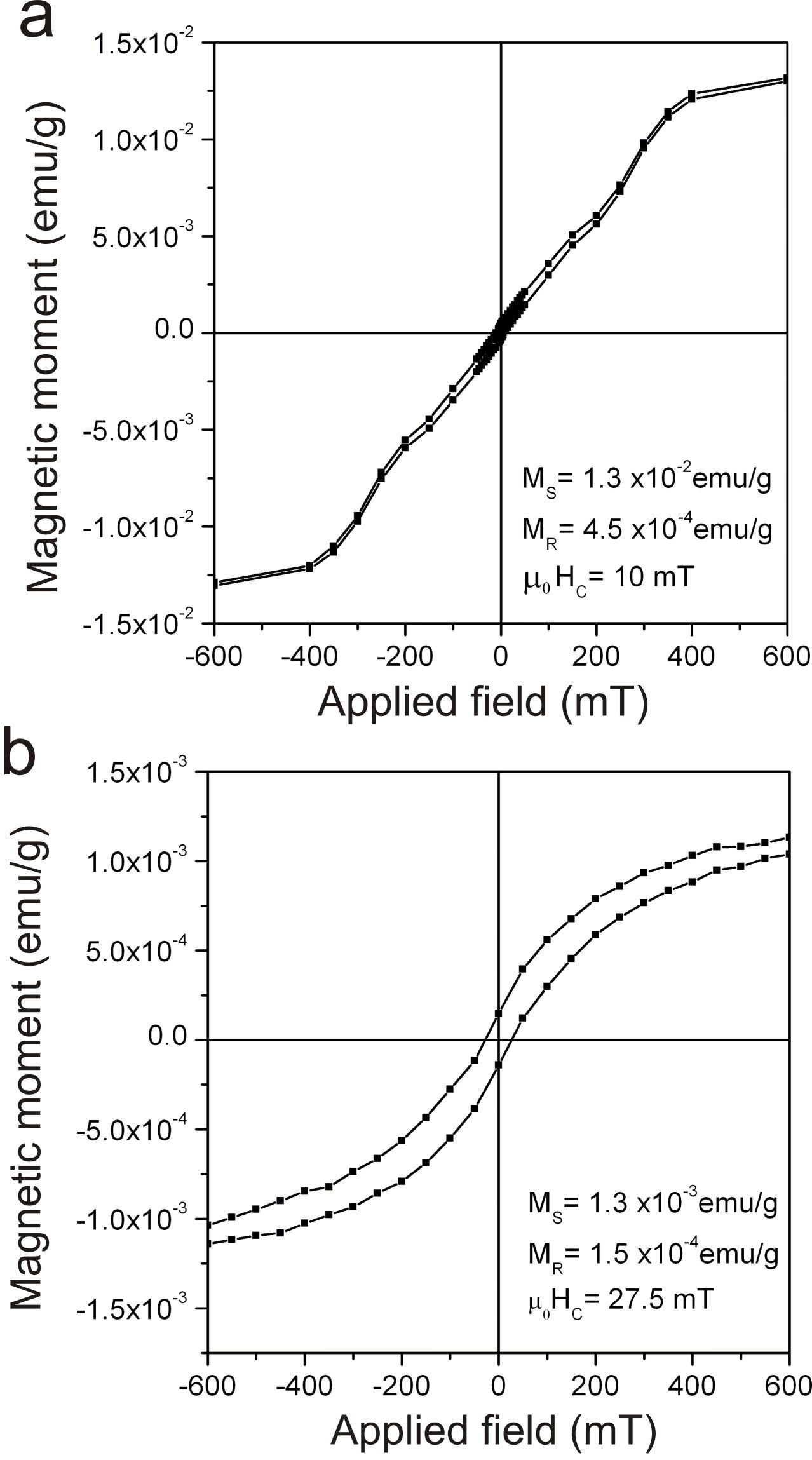}
   \caption{Out-of-plane ($H~\parallel~c$) SQUID magnetization measurement on HOPG after substraction of the diamagnetic signal at 5~K (a) and 300~K (b). The diamagnetic background signals were $\chi=-3.2\times10^{-4}$~emu/g~mT at 5~K and $\chi=-1.0\times10^{-4}$~emu/g~mT at 300~K.}
   \label{fig2}
\end{figure}

The bulk magnetization of the HOPG samples have been analyzed with SQUID magnetometer at 5~K and 300~K.  In figure~\ref{fig2}, out-of-plane magnetization loops of HOPG after substraction of linear diamagnetic background signals are shown. They demonstrate small hysteresis even at room temperature, which is a clear sign of ferromagnetic order. The out-of-plane saturation magnetization reaches value $0.013$~emu/g at 5~K. In-plane magnetization loops have been measured on the same sample as well. They have shown ferromagnetic hysteresis loops comparable to SQUID measurements on HOPG reported by P. Esquinazi et al.~\cite{Esquinazi2}. The in-plane magnetization loops saturated at one order smaller values $2\times10^{-3}$~emu/g than in the out-of-plane configuration at 5~K. The coercive field and remnant magnetization have been found similar in both in-plane and out-of-plane magnetization measurements. In the work of P. Esquinazi at al.~\cite{Esquinazi2}, the ferromagnetic signals were measured up to temperature 500~K.

The observed high temperature ferromagnetism in HOPG can have different possible origins. The first one is obviously ferromagnetism due to magnetic impurities. HOPG samples, as it has been studied previously~\cite{Esquinazi2}, contain small fraction of magnetic elements. Therefore, we have analyzed the HOPG samples for impurity concentration by particle induced X-ray emission (PIXE) in the bulk material and by low energy ion scattering (LEIS) at the surface. As a main magnetic impurity in PIXE was found Fe with concentration around 20~$\mu$g/g and Ti with concentration 4~$\mu$g/g. Other magnetic and metallic impurities have been found below 1~$\mu$g/g. The surface analysis by LEIS have not detected any magnetic elements indicating that the concentration of all these elements is below 100~ppm. The measured content of Fe impurities in HOPG is not sufficient to produce the ferromagnetic signal shown in figure~\ref{fig2}. The amount of 1~$\mu$g/g of Fe would contribute maximally $2.2\times10^{-4}$~emu/g to the magnetization and for Fe or Fe$_3$O$_4$ clusters, the magnetic signals would be even smaller~\cite{Esquinazi2}.

Another possible source of the shown up ferromagnetic behavior are the defect structures in graphite. Line defects occur naturally in graphite as edges and grain boundaries. Graphite edges have been extensively studied both theoretically \cite{Fujita,Nakada} and experimentally \cite{Kobayashi1,Niimi,Kobayashi2}. There are two typical shapes for graphite edges: armchair and zigzag. Only zigzag edges are expected to give rise to the magnetic ordering due to the existence of the edge state \cite{Fujita}. STM experimental results on step edges of graphite, however, showed that zigzag edges are much smaller in length ($\approx$ 2~nm) than those of armchair edges and less frequently observed \cite{Kobayashi2}. At the same time, this would also suggest the occurrence of a one-dimensional ferromagnet of micrometer size at room temperature. We rather believe that step edges are created on HOPG at places where bulk grain boundaries cross the surface. During the cleavage of the graphite crystal these grain boundaries are the weakest points of the graphite crystal. A step edge created by this way would have the same orientation and geometry as a grain boundary underneath.

\begin{figure}[t] 
   \centering
   \includegraphics[width=3.3in]{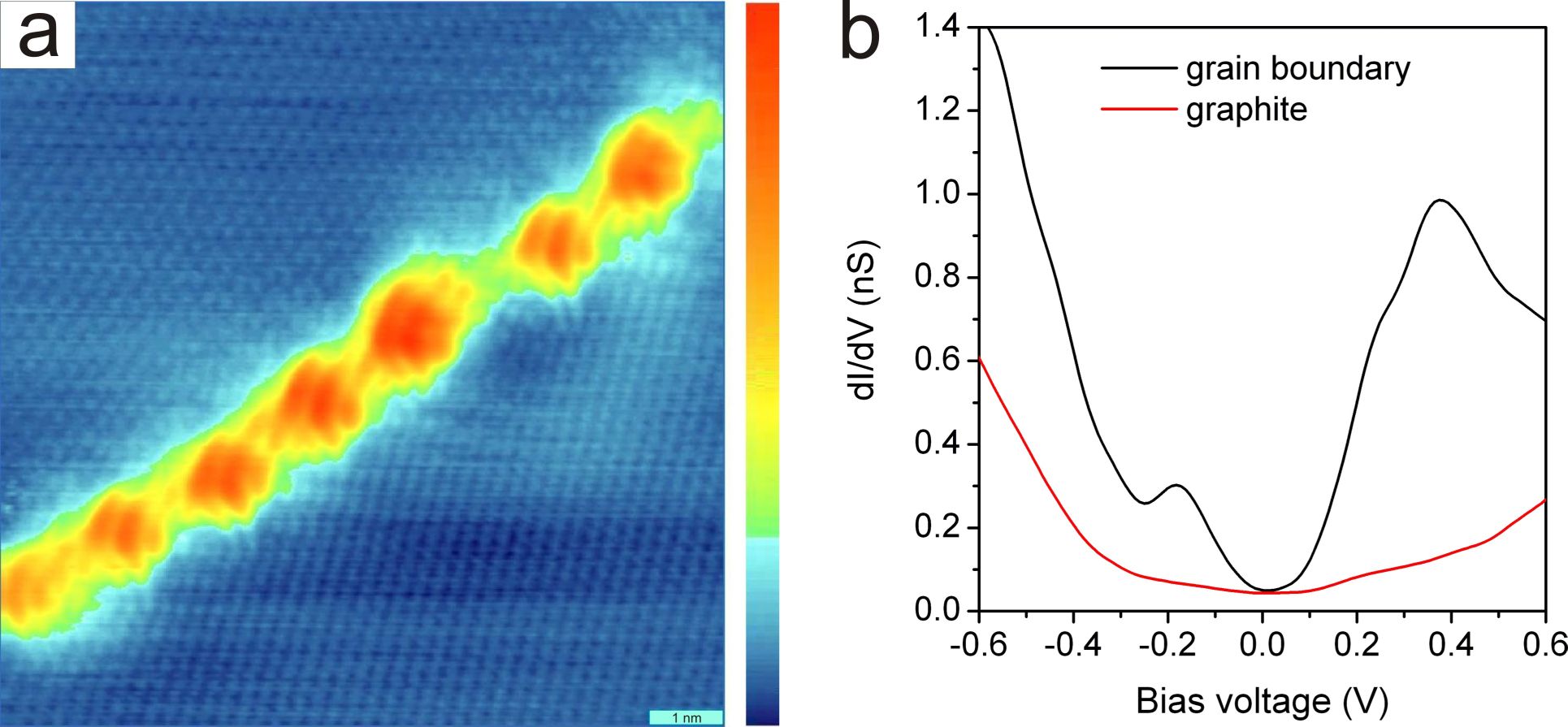}
   \caption{(Color online) (Color online) (a) STM image of a grain boundary on HOPG showing a 1D superlattice with periodicity $D=1.4$~nm. The angle between two graphite grains is 18$^\circ$ and the angle between the grain boundary and graphite lattice is 9$^\circ$. Scanning parameters: $10\times10$~nm$^2$, $U=0.6$~V, $I_t=0.4$~nA. (b) STS on the grain boundary and on the bare graphite surface (tunneling resistance 0.9~G$\Omega)$. The grain boundary shows localized states at -0.18~V and 0.36~V.}
   \label{fig3}
\end{figure}

Grain boundaries in HOPG have been studied in great detail by AFM and STM \cite{Cervenka1,Cervenka2}. Grain boundaries are inevitable defects in graphite because of polycrystalline character of the crystal. They are formed between two rotated grains during the crystal growth. They extend over step edges and form a~continuous network all over the graphite surface. Grain boundaries show a~small or no apparent height in AFM (see figure \ref{fig1}). On the other hand, grain boundaries exhibit a very distinct sign in STM, where they appear as a one dimensional superlattice with the height corrugation up to 1.5~nm. An example of STM and STS on a grain boundary with the periodicity $D=1.4$~nm is displayed in figure \ref{fig3}. Since grain boundaries have almost no height in AFM and the corrugation of STM is given by convolution of the topography and the local density of states (DOS) of the substrate, hence grain boundaries must be endowed with an enhanced charge density compared to the bare graphite surface. Apparently this is seen in the STS of a grain boundary (figure~\ref{fig3}b), which in addition shows two strong localized states at~-0.18~V and 0.36~V. These states are not observed on the bare graphite surface. Similarly, two localized states have been measured for different grain boundaries~\cite{Cervenka2}.

Defects in graphene break the electron-hole symmetry, which lead to creation of localized state at the Fermi energy and to the phenomenon of self-doping, where charge is transferred to/from defects to the bulk~\cite{Vozmediano,Peres1}. The self-doping of defects is in accordance with our experimental observations showing enhanced charge density at the grain boundaries. Since graphene systems have low electron densities at the Fermi energy, electron-electron interactions play an important role as the recent experiments showed~\cite{Peres1,Gruneis}. In the presence local repulsive electron-electron interaction the localized states will become polarized, leading to the formation of local moments~\cite{Vozmediano}. This has been illustrated in DFT studies of point defects in graphite such as vacancies and hydrogen-terminated vacancies. These defects revealed to be magnetic having a local magnetic moment larger than 1$\mu_B$~\cite{Lehtinen,Yazyev}. In graphene, the indirect RKKY interaction mediated via valence electrons between these local moments has been found to be always ferromagnetic due to the semimetallic properties of graphene \cite{Vozmediano}. The presence of ferromagnetism has been studied theoretically in the phase diagram of pure and doped graphene \cite{Peres2}. It has been shown that ferromagnetic order can be stabilized at low doping when the exchange coupling is sufficiently large and as well at higher doping due to a disorder which further stabilizes the ferromagnetic phase. In graphite, however, because of finite density of states at the Fermi energy there is expected to be competing ferro- and antiferromagnetic coupling between the moments. This gives an additional oscillating term with the oscillation period determined by the Fermi momentum of electrons (holes) \cite{Dugaev,Vozmediano}. In the DFT study of an 3D array of single vacancies in graphite, different supercells containing single vacancy have been studied~\cite{Faccio}. Ferrimagnetic order has been supported up to the distance 1 nm among the vacancies, while $5\times5\times1$ supercell (1.23 nm separated vacancies) did not show a net magnetic moment in graphite~\cite{Faccio}. In graphene, the $5\times5$ supercell exhibited still a net magnetic moment of 1.72$\mu_B$~\cite{Faccio}.

In a similar way, grain boundary in one layer of graphite can be visualized as a one-dimensional line of equidistantly distributed defects, where the superlattice periodicity gives the distance between the defect sites. The defects in grain boundaries are not single vacancies, for which a trigonal symmetry would be expected to be observed in STM, but rather more complicated defects. If the spin polarized electron states were created in a grain boundary, the exchange splitting would be in order of 0.6~eV in our experiment (figure \ref{fig3}b). We have observed two split localized states around the Fermi energy at different grain boundaries with periodicities below 4~nm. Similarly like in the DFT study of an 3D array of single vacancies in graphite~\cite{Faccio}. Another supporting evidence that the ferromagnetism originates from grain boundaries is the fact that grain boundaries and step edges are the only defects observed in STM experiments on graphite surface. No point defects have been detected on the graphite surface with STM. Secondly important aspect is the two dimensional character of the grain boundaries, which explains most of the features from MFM and SQUID measurements.

We assume that grain boundaries are propagating along the \textit{c}-axis of the graphite crystal creating 2D plane of defects. The distance between the defects in grain boundaries is determined by the superlattice periodicity in the plane and by the interlayer separation 0.33~nm along the \textit{c}-axis. As it was described before, step edges can be the manifestation of the grain boundaries buried underneath them. The ferromagnetic signal would then come from 2D grain boundary planes formed through the bulk crystal. Moreover, an infinitely extended 2D magnetic plane with in-plane magnetization is stray-field-free and therefore it can exist in the single-domain state \cite{Allenspach}. Accordingly, in-plane magnetized grain boundary plane should show a single magnetic domain, which supports the observation of only one magnetization direction in MFM measurements. Due to crossings among grain boundaries, minimum energy configuration would lead to magnetization aiming in the \textit{c}-axis of HOPG. 2D character of grain boundaries is also in accordance with the higher saturation magnetization measured along the \textit{c}-axis than along the in-plane direction. It is expected that most of the grain boundaries have small tilt towards the c-axis. Therefore, a larger magnetic field will be needed to align the local magnetic moments along the \textit{c}-axis than in the graphene planes, where the magnetic axis stays in the 2D grain boundary plane.

The electrons involved in the ferromagnetic behavior in graphite are $sp$ electrons and therefore the well known theory of magnetism based upon the unfilled character of $3d$ or $4f$ electrons energy levels cannot be directly applied. In the proton irradiated carbon studied by x-ray magnetic circular dichroism, it was demonstrated that the magnetic order originated only from the carbon $\pi$-electron system \cite{Ohldag}. Similarly it has been found by Faccio et al. \cite{Faccio} that $\pi$ electrons are responsible for the magnetism in the 3D array of single vacancies.

Ferromagnetism with high Courie temperature $T_C$ of itinerant $sp$ electrons in narrow impurity (defect) bands has been recently studied in theory. This theory has been applied to CaB$_6$ but it is well applicable to graphite with narrow grain boundary states too. It has been argued that Stoner ferromagnetism with high Curie temperatures $T_C$ can be expected for $sp$ electron systems with narrow impurity states~\cite{Edwards}. In this theory, correlation effects do not reduce the effective interaction which enters the Stoner criterion in the same way as in a bulk band ferromagnets. Moreover, the spin wave excitations may not be effective if full spin alignment is maintained in lowering $T_C$. The value of $T_C$ can be thus close to the value given by Stoner theory unlike for other bulk ferromagnetic metals. According to the above analysis, we believe that the grain boundaries are playing a crucial role in the observed ferromagnetism in HOPG.

In conclusion, ferromagnetic signal has been observed in HOPG by magnetic force microscopy at room temperature. The observed ferromagnetism has been attributed to originate from unpaired electron spins localized at defects sites of grain boundaries. Scanning tunneling microscopy (STM) and spectroscopy (STS) have revealed localized states and enhanced charge density of grain boundaries.

\acknowledgments
The authors are grateful to R. Lavrijsen and B. Koopmans for SQUID measurements, P. H. A. Mutsaers for PIXE analysis and H. H. Brongersma for LEIS measurements. This research was supported by Nanoned.

\end{document}